\documentclass{svproc}
\usepackage{url}

\usepackage{graphicx}
\usepackage{cite}
\usepackage{makecell}
\usepackage{multicol}
\usepackage{footmisc}

\begin{document}
\mainmatter              
\title{Pre-Release Experimentation in Indie Game Development: An Interview Survey}
\titlerunning{Pre-Release Experimentation in Indie Game Development}  

\author{Johan Linåker\inst{1} \and 
Elizabeth Bjarnason\inst{1} \and
Fabian Fagerholm \inst{2}}
\authorrunning{Linåker et al.}
%

\institute{Lund University, Dept. of Computer Science, Lund, Sweden\\ \email{\{johan.linaker | elizabeth.bjarnason\}@cs.lth.se} \and 
Aalto University, Helsinki, Finland \\
\email{\{fabian.fagerholm\}@aalto.fi}}

\maketitle              
\begin{abstract}
[Background] The game industry faces fierce competition and games are developed on short deadlines and tight budgets. Continuously testing and experimenting with new ideas and features is essential in validating and guiding development toward market viability and success. Such continuous experimentation (CE) requires user data, which is often limited in early development stages. This challenge is further exacerbated for independent (indie) game companies with limited resources. [Aim] We wanted to gain insights into CE practices in pre-release indie game development. [Method] We performed an exploratory interview survey with 10 indie game developers from different companies and synthesised findings through an iterative coding process. [Results] We present a CE framework for game development that highlights key parts to consider when planning and implementing an experiment and note that pre-release experimentation is centred on qualitative data. Time and resource constraints impose limits on the type and extent of experimentation and playtesting that indie companies can perform, e.g. due to limited access to participants, biases and representativeness of the target audience. [Conclusions] Our results outline challenges and practices for conducting experiments with limited user data in early stages of indie game development, and may be of value also for larger game companies, and for software intensive organisations in other industries. 

\keywords{Game development, Continuous Experimentation, User Research, User data, Indie Game Company}
\end{abstract}
\section{Introduction}
The game industry is ever growing with increasing yearly turnarounds and number of employees. For example, the Swedish game industry represents 4.1 percent of the country's national gross service export with a yearly turnaround  of 86.5 billion SEK (including international subsidiaries) and a workforce of around 25,000 people globally, including 8,500 employees in 939 companies located in Sweden ~\cite{swedishgamesindustry2023developerindex}. The industry is fiercely competitive, and game development companies are pressured by short deadlines and need to stay cost-efficient\cite{schmalz2014risk}. Therefore, betting on a wrong idea or a feature that is not technically or commercially viable poses a significant business risk.

In the broader industry, software-intensive organisations have started to adopt Continuous Experimentation (CE) practices~\cite{ros2018continuous} to mitigate the risk of releasing products with weak market viability. An experiment-driven development approach is then used to continuously evaluate new features and qualities based on user feedback~\cite{fagerholm2017right}, and hypotheses are defined upfront based on the expected outcome in terms of technical and/or commercial viability~\cite{auer2021controlled}. 

Experimentation would ideally be performed throughout game development from ideation and pre-release to post-launch, and thereby guide development and product planning in producing an engaging game experience that maximises the value creation~\cite{yaman2018continuous}. The stages prior to releasing a game are extra critical as this is where the majority of the development occurs. As development progresses, the cost and impact of pivoting increase.

CE requires access to user data to test hypotheses~\cite{fagerholm2017right}, which is especially challenging prior to release of a game. Since the number of  users is naturally limited at this point, so is the amount of available user data and feedback~\cite{yaman2018continuous}. This challenge is especially taxing for smaller game development companies with limited resources and without the backing of larger game publishers. These companies, also referred to as independent (\textit{indie}) game companies are commonly small and new companies with limited resources, experience and practice comparable to established game companies and studios~\cite{grabarczyk2016every}.  

There is limited research of CE practices in the game industry, and specifically for indie game studies ~\cite{engstrom2020game}, with some notable exceptions~\cite{yaman2018continuous, edison2023experimentation}. In this study, we aim to address this gap by investigating \textit{how CE may be applied in the context of pre-release indie game development}. We conducted an exploratory semi-structured interview survey with 10 game developers from different indie game companies. Findings are synthesised into an emerging framework of continuous experimentation in game development that highlights five key parts to consider when planning and implementing experimentation in game development. In the future, we aim to further validate and refine our initial framework through case studies and expand our research to include larger game companies.

\section{Related Work}

Continuous Experimentation (CE) is a software development approach where product design decisions are based on results from field experiments with real users~\cite{fagerholm2017right, auer2021controlled}. Several methods and frameworks have been proposed for CE, involving similar activities: identifying and articulating product assumptions, turning these into testable hypotheses, running experiments to validate the hypotheses, and interpreting the experiment results to determine whether the assumption held or not~\cite{holmstromolsson2014hypex, fagerholm2017right}. This allows the product organisation to iteratively find designs that promote desired user behaviours, such as increased purchases.

Many CE-like activities can be observed in game development.
Game analytics is a common practice that has been used for, e.g., in-game balancing, identifying bottlenecks in game levels, detecting bugs, reducing costs and risks, and negotiating with investors and publishers~\cite{su2021gameanalytics}.
Its use of field data to drive decision-making is common with CE.
Some specific aspects of CE have received direct attention in game development~\cite{engstrom2020game}, e.g., the exploratory use of prototyping in the requirements engineering process~\cite{bjarnason2023empirically}, and monetary feature-optimisation in post-deployment.
The area of User research also provides some input into how user feedback can guide game development~\cite{pagulayan2018applied}.
User research shares with CE the overall idea of basing development on an empirical approach.

Research describes a need to match experimentation practices to the different stages of game development~\cite{aleem2016game}.
Qualitative methods are primarily used in early development stages when the amount of quantitative user data is limited~\cite{yaman2018continuous}.
Play-testing is highlighted as the main overarching method, while A/B testing is common in the later post-production stages~\cite{hyrynsalmi2018minimum}.
Identifying and isolating features for evaluation is a challenge~\cite{koskenvoima2015small}, especially in the early development stages~\cite{jarvi2015lean}.
The feature set needs to be holistic enough to not loose important aspects required to enable game-play, while still enabling the experiment to evaluate defined Key Performance Indicators (KPIs)~\cite{hyrynsalmi2018minimum}.

A distinction can also be noted in relation to the game genre and context.
For example, in the mobile game context and freemium-type games~\cite{koskenvoima2015small}, there is a strong belief in quantitative methods.
Andersen et al., demonstrate the potential of quantitative experimentation in guiding feature development in gaming~\cite{andersen2011on}. However, many game companies also rely on unstructured approaches.

Depending on the purpose and aspects being experimented on, different metrics are required, such as game mechanics (the rules of the game), game dynamics (what happens when the game is played), and aesthetics (the feeling of fun experienced by the player)~\cite{jarvi2015lean}.
When designing an experiment, each aspect requires specific consideration regarding the technical implementation and commercial viability~\cite{rosenfield2017mvps}.
Validation of aesthetics, or fun factor (regardless of technical or commercial viability) is specifically highlighted as a key challenge, due to the need to connect with a relevant target audience~\cite{schmalz2014risk}, while limiting the risk of leaking differentiating ideas and details to the public and potential competitors~\cite{aleem2016game}.
Therefore, play-testing is usually performed with internal developers or trusted associates, despite the risk of cognitive and confirmation biases~\cite{yaman2018continuous}.
Chueca et al. points to the need for new methods, tools, and processes to identify and validate requirements relating to the user experience of games~\cite{chueca2023consolidation}.

Regardless of the many investigated aspects, the full CE cycle in game development has to date received limited attention with only recent work starting to emerge~\cite{yaman2018continuous, edison2023experimentation}.
Scientific studies on holistic experimentation frameworks and processes are still missing despite the competitive and high-paced nature of the gaming industry, which points to a need for more research in this context on continuous and iterative experimentation.
Specifically, the early stages of game development are highlighted as critical since many game ideas never enter the production stage~\cite{schmalz2014risk}.
Thus, this is an area for future research~\cite{aleem2016game, chueca2023consolidation}. 

\section{Research Design}

An interview survey \cite{MethodRef}
of how CE may be applied within indie game development prior to releasing a game has been performed. We sampled interviewees with the goal of gaining a wide understanding of practice across different indie game companies. One representative per indie company was deemed sufficient to gain an understanding of the practice and experience at these very small companies. Representatives from ten indie company were identified at an indie game conference based on informal conversations from which initial observations were collected in field notes. These company representatives were then interviewed online, see Table \ref{tab:interviewees} for an overview of  interviewees I1-I10. 

\begin{table}[!b]
    \centering
    \caption{Overview of interviewees (years of experience, current role) and their indie company (number of employees, game genre and platform, geographic location.}
    \label{tab:interviewees}
    \begin{tabular}{|p{0.5cm}|p{0.9cm}|p{3.5cm}|p{0.9cm}|p{2.8cm}|p{1.8cm}|p{1cm}|}
    \hline
    
        \makecell{ID} & 
        \makecell{Exp. \\ (Yrs)} & 
        \makecell{Role(s)} & 
        \makecell{Nbr \\ Empl.} & 
        \makecell{Game genre(s)} & 
        \makecell{Platform(s)} & 
        \makecell{Main \\ loc.} \\
    \hline
        I1 &  
        7 &  
        Producer, Co-Founder &  
        4 &  
        First-person-shooter & 
        Desktop & 
        CH \\
        
         I2 & 
         18 & 
         Lead developer, Founder & 
         1 &
         Open world, Strategy & 
         Desktop &
         SE, US \\
         
         I3 &  
        13 &  
        CEO, Lead developer and Designer, Founder &  
        3 &  
        AI-based Role play & 
        Desktop & 
        SE \\
         
         I4  &  
        2 &  
        Director, Lead developer, Co-Founder & 
        8 &   
        3D vehicle combat action roguelite & 
        Desktop & 
        SE \\
         
         I5  &  
        6 &  
        Lead designer, Co-Founder &  
        2,5 &  
        3D rougelite & 
        Desktop & 
        NL \\
         
         I6 &  
        5 &  
        Game director, Co-Founder &  
        9 &  
        3D Role Play & 
        Desktop, console & 
        SE \\
         
         I7 &  
        14 &  
        CEO, Co-Founder &  
        50+ &  
        Puzzle, Brawler, Action & 
        Mobile & 
        UA \\
         
         I8 &  
        4 &  
        Development and design lead, Co-Founder &  
        2 &  
        Cozy wholesome games & 
        Desktop & 
        SE \\
         
         I9 &  
        5 &  
        Lead Developer, Founder &  
        1.5 &  
        Platform, action & 
        Desktop & 
        NL \\
         
         I10 &  
        5 &  
        CEO, Co-Founder&  
        2.5 &  
        Sustainability-focused cozy games& 
        Mobile & 
        SE \\
         
    \hline
    \end{tabular}
\end{table}

Each interview lasted for about 30 minutes. I1-I6 were interviewed by the first and second author together, while I7-I10 were interviewed by the first author only. The interviews were semi-structured using an interview questionnaire ~\cite{SupplementaryMaterial} focused on open questions regarding the use and experience of CE and related practices such as game user research, testing, and requirements engineering. Review of literature provided authors with an initial understanding of the problem domain, enabling probing questions during interviews.

The interviews were recorded, transcribed with an offline instance of WhisperX, and analysed using open and axial coding techniques~\cite{saldana2021coding}. The 1st and 2nd author coded I1 and I2 separately, and their two code books were merged and disagreements settled. For the remaining interviews the two authors took turns in coding and peer-reviewing each interview, and settled disagreements. 

The coding rendered ten high level themes, in addition to demographics. These themes were informed inductively from the interview data and from our knowledge of related work. The final code book can be found in the supplementary material~\cite{SupplementaryMaterial}. When coding I6, we  noted some saturation among the findings and first-level codes introduced. By I10, the data was mainly of a confirming nature, adding only nuances to existing first-level codes. 

Steps were taken to address \textbf{threats to validity} of the study. An audit trail was maintained to retain traceability throughout data collection, analysis, and reporting using references to I1-I10. Inclusion of quotes has been prioritised to provide richness and context in the reporting, and to enable anecdotal generalisation based on the data. Each interviewee was provided with an early version of this article to enable member-checking and validation of our synthesised findings. Collaborative and peer-review coding together with peer-debriefing was used to enhance construct validity and reduce the risk of researcher bias.

\section{Results: Continuous Experimentation Framework}

\begin{figure}
    \centering
    \includegraphics[width=1\linewidth]{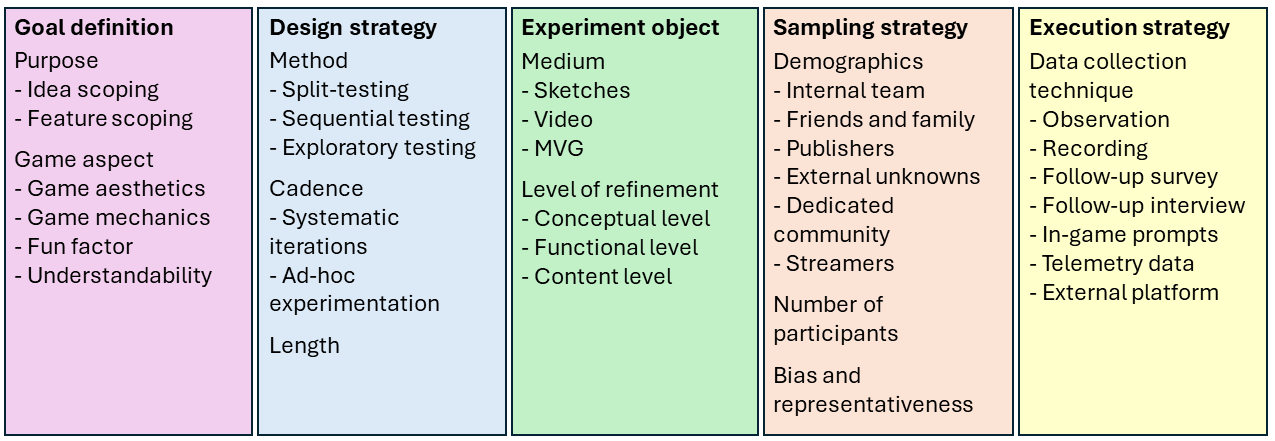}
    \caption{Overview of our framework for CE in game development including key parts of experimentation in the early stages of game development.}
    \label{fig:model}
\end{figure}

Our results are synthesised into a framework for CE in game development, see Fig.~\ref{fig:model}. The framework consists of five main parts involved in experimentation, namely \textit{goal definition}, \textit{design strategy}, \textit{experiment object}, \textit{sampling strategy}, and \textit{execution strategy}. Each part is described based on our empirical data and with quotes derived from our interviewees (I1-I10, see Table \ref{tab:interviewees}.)

\subsection{Goal Definition} 
We find that the goal of experimentation consists of a combination of the \textit{purpose} of an experiment and the \textit{game aspect} being evaluated. In the very early stages of development, the \textbf{purpose} primarily concerns the scoping of the overarching \textit{game idea}, while it later shifts to \textit{feature scoping}. Selecting and \textbf{scoping the game idea} is critical since it determines the future direction of the development and, ultimately, the success of the game. This scoping affects the development team, the target audience, and potential stakeholders, including publishers. For example, I6 described experimenting with three game ideas by presenting conceptual sketches of these to a potential publisher and using this feedback to select which idea to pursue and to invest further development effort into. When experimenting for the purpose of \textbf{feature scoping} \textit{``the decision on what to include in the game''} (I2) is based on experimentation results, both regarding which features to include and the detailed scoping of these.

Experiments often focus on validating one or more \textbf{game aspects}, and four such aspects emerged through our interview survey, namely \textit{game aesthetics}, \textit{game mechanics}, \textit{fun factor}, and \textit{understandability}. Experimentation is performed for these aspects, both with the purpose of idea scoping and feature scoping. An experiment that focuses on \textbf{game aesthetics} considers the graphical style and artwork of the game. While the final artwork is conceived and evaluated in the later stages of pre-release (Alpha and Beta), we found some instances where the aesthetics are considered already in the idea-generation stage. For example, I10 described experimentation using early sketches that were shown to the target audience before any playable demo was created. Our interviewees described that the goal when experimenting with game aesthetics is to evaluate the conceptual look and feel of the game rather than the exact artwork to be implemented in the final product. For example, I9 described using placeholder-level art created by an AI-based image generator (Midjourney AI) to experiment and get the right \textit{feel} early on, and that this cost-effective means of experimentation enabled creating a playable demo after only one month. 

In the early stages of development, there is a strong focus on experimentation with the \textbf{game mechanics}, which concerns the dynamics, features, and physical mechanics as experienced when playing a game. An example comes from I1, who is developing a first-person-shooter game where the main character is a non-human who neither moves or acts like a human, nor has hands to hold a gun with. In this case, early and frequent experimentation was performed to develop \textit{``the core mechanics... running around, shooting, reloading''} (I1) for this new type of character. The main goal of this experimentation was to figure out how these new mechanics should work to provide the right feel and game experience. Once this was in place, the team moved on to content production and working towards a beta release. Similarly, another interviewee described focusing on \textit{``mechanics like weapons and items... to see what kinds of choices players were making and if they felt that those were satisfying''} (I4).

The mechanics and aesthetics of a game tie into the softer aspect of the \textbf{fun factor}, which regards the user experience, feel, and fun experienced by the user, i.e., the \textit{game user experience}~\cite{stahlke2018usertesting}. I1 highlights how this aspect can, and should, be evaluated early on in development, even before the game has nice-looking aesthetics. \textit{``Players would have fun in a level that is pretty much a grey box. There is actually not much in there, but you still have your features... That is basically the best position. If you can test it so far that players enjoy it and actually have fun, even though the level is still ugly as hell''} (I1). This experimentation with game mechanics to evaluate fun factor should ideally continue throughout the development. I4 describes first focusing on evaluating players' game user experience and how they react to the core loop and mechanics of a game, and later shifting to experiment with \textit{``different areas of the mechanics, [e.g.,] its weapons and items. We really wanted to see what kind of choices players were making and if they felt that those were satisfying''} (I4).

The \textbf{understandability} of a game is an aspect that is closely related to \textit{fun factor}, and concerns how easily players comprehend and understand the logic of a game. The complexity of this aspect varies depending on the type of game. For example, for a strategy game, the player needs to \textit{``learn... and understand the basic game mechanisms''} (I2) to play and enjoy the game and thus has a steeper learning curve compared to a first-shooter game. In this case, I2 performed experimentation with a tutorial to validate the learning aspects and to see how well players understood and learnt to use the game mechanisms.

\subsection{Design strategy} 

Our interviewees describe a range of strategies used when designing and setting up experiments that vary regarding the kind of \textit{method} used, and the \textit{cadence} and \textit{length} of experiments. The described design strategies vary regarding formality and scope, and depend on the experimentation goals, type of game, current development stage, and available resources. 

We observe three main \textbf{experimentation methods} in our material, namely \textit{split}, \textit{sequential}, and \textit{exploratory testing} that represent strategies for exploring different options. Either, two versions are compared (split), options are exploring one by one (sequential), or the design space is freely explored (exploratory). \textbf{Split testing} is generally considered to be expensive in terms of resources, which is why this was mostly described for the early ideation and prototyping parts of pre-release. Split testing is usually limited to two versions (i.e., A/B testing). Several interviewees (I3, I5, I6) described experimenting by implementing two and three early alternative versions of their games, ranging from playable demos to conceptual art, and presenting these to potential publishers. The obtained feedback was often used to guide the indie game companies decisions on which alternative to pursue. However, I3 described that the experiment was inconclusive due to conflicting feedback from different publishers, and in the end the company decided to go with the game that had the lowest development cost. 

In later stages of the pre-release, split-testing often becomes more focused on specific features. In this stage, the cost of implementation is still perceived as low, or reasonable in relation to the perceived benefit. I3 describes developing two versions of their main game, one of which contained more advanced AI functionality. The participants of the experiment selected which version they preferred to play and feedback was collected through a survey. Based on 60-70 survey responses the decision was to adopt the new, more advanced functionality. I1 also described using multiple prototype variants to experiment with in the early parts of pre-release. 
I7 stands out in our survey by describing comparatively mature experimentation practices using A/B-tests. The interviewee's experience of this is from development assignments for larger mobile game development companies. This includes the case where \textit{``you have [game] mechanics that you are not sure will work for your game... so we develop two different versions in which those mechanics could be done in two different ways. Our team generates a hypothesis that [the new functionality] will improve for the specific KPI, for example, the play time''} (I7). 

\textbf{Sequential testing} is an experimentation method where the option considered to have the best potential is evaluated first, and if the feedback is negative, the alternative solution is implemented and evaluated. Our interviewees generally prefer this method when development matures beyond the early ideation and prototyping stages. One interviewee described focusing their experimentation on functionality that is \textit{``not working, and then we try to iterate on it, and then see if it has improved. But, it is not really A/B testing. We just test the new version and see if the reactions have improved... Most of the time, it is an improvement''} (I5). I6 exemplifies a more comprehensive form of sequential testing where they  experimented with alternative artwork and game mechanics compared to the main game. Based on the obtained feedback, the ideas considered successful were then re-implemented in the main game. Another example of sequential testing was described by I9, where they had experimented by replacing a gun with an alternative mechanisms to evaluate the hypothesis that the gun was being used too repetitively, impacting negatively on the user experience. In this case, the experiment revealed that the users enjoyed, and preferred, the gun, which led to re-implementing it.

\textbf{Exploratory testing} is a less formal experimentation method where there is no predefined hypothesis but rather a set of questions or an open mind when entering an experiment, e.g. when playtesting. The goal of the experiment can be general, or aimed at considering specific aspects such as game mechanics or fun factor. I5 explains it as: \textit{``Sometimes, we go in with a questionnaire, sometimes we just have some questions at the back of our minds while observing, and sometimes we are going in [with a] fully open [mind]... If we have something that we know we want to know more about, then it is really focused... if it is more ...how does the game play experience feel ... then it is more open.''} Also, I4 describes typically having a set of predefined questions when entering a playtest. Lessons and takeaways help inform the backlog and prioritise feature development and bug fixing. I10 describes maintaining an open mind in playtests focusing on game mechanics and fun factor, and then observing how players move around and interact with the game environment. As stated by I6, \textit{``fun is so subjective''} and the evaluation of fun factor requires more than pure quantitative data. 

The \textbf{cadence} with which experiments are perofmred varies through \textit{structured systematic experimentation} to \textit{ad-hoc experimentation}. I1 describes using systematic experimentation in early prototyping stages through small and quick iterations. For this, a small set of easily available participants from within their co-working space are used as playtesters. In later stages, more development and refinement are conducted between tests, adding longer periods in between experiments. I2 describes a similar approach, but notes that he sometimes mixes the cadence, e.g. \textit{``if I do one month of rapid play testing, I might afterwards pause for a bit to take more long-term decisions, and introduce larger feature implementations, followed again by more intense testing''}. Several interviewees (I1, I4, I6, I9) note how the time and resources required to plan, execute, and analyse the playtests are costly and limited in comparison to what larger AAA game development companies can perform. Others, such as I5 and I8, describe a more \textit{ad-hoc experimentation}, e.g., dependent on the availability of local and externally hosted playtesting sessions. 

The \textbf{length} of the experiments is described to range between a weekend to a week by the majority of our interviewees. I4 describes how experiments may be longer, spanning up to several weeks, during which the game is continuously updated, e.g. with corrections. The consequences of experimenting with a constantly changing version of the game is that the \textit{``feedback that you get at the beginning of the playtest and the feedback that you get at the ending of the playtest is actually feedback on two different builds''} (I4).

\subsection{Experiment Object}

The experiment object refers to the artefact that is evaluated in an experiment. The \textit{medium} (sketches, video, or MVG) and \textit{level of refinement} (conceptual, functional, or content level) of the experiment object need to be aligned to the goals and the design strategy of the experiment. Game developers also need to consider the time and resources required to produce the experiment object, which also relates to the current stage of the game development. For the \textbf{medium} in the initial stages, interviewees describe using \textit{sketches} and conceptual designs of the art and imagined gameplay to evaluate the overall game idea. \textit{Videos} demonstrating gameplay may be an option to communicate an idea quickly, while playable \textit{Minimum Viable Games (MVGs)} offer a direct experience for the participants of the experiment. Even if these \textit{``early versions of the game ... aren't very refined, they work''} (I5) and allow game developers to get early feedback on their ideas which \textit{``informs our backlog''} (I4) and is thus fed directly back into the game design and development. For this reason, experimenting with an MVG is often preferred even if the MVG is \textit{``very unpolished''} (I2) and requires more time and effort to produce than prototyping a new idea in isolation. One interviewee motivated this preference with a desire to \textit{``get it [the change] into the game as quickly as possible ... to see if it works with the rest''} (I2).

We have identified three main \textbf{levels of refinement} of the experiment object with increasing complexity and maturity, namely \textit{conceptual}, \textit{functional}, and \textit{content level}. On the \textbf{conceptual level}, high-level ideas and visions are evaluated using, e.g., sketches, videos of gameplay, and rough prototypes of the art work and envisioned gameplay. For example, I6 described using three conceptual designs to present their game ideas to publishers early one, while I10 described presenting various art designs to the intended target audience. I4 described using non-executable (abstract) experimentation objects, such as design and conceptual art and descriptions, to evaluate ideas and design options internally within the company. When experimenting at the \textbf{functional level}, the core mechanics and loop of the game are demonstrated. I1 describes how they \textit{``focused on getting the core mechanics of the game, the core loop right, which basically meant ... running around, shooting needles, reloading, tapping at the right time... that's basically the core loop... [We] would look deeply into things like how much head bobbing, ... [should there be] when [moving side-ways] left or right?''}. I4 described how the length of an experiment initially was limited to a two-week iteration to present a functional prototype, followed by a longer 2-3 month iteration where an early playable demo was created and evaluated by friends and family. On the \textbf{content level}, art work and level design is added and finalised, but not necessarily at a very polished level as hinted by I4. One interviewee differentiates between the functional and content levels as \textit{``...you need the feature side, which is all the logic, which is design dictating what code to build, and whether it works, whether it's a proof of fun ... And, the content side is stretching out that fun, putting these features in varied situations, [i.e.,] level design''} (I1).

\subsection{Sampling strategy} 










The people that participate in an experiment are selected, or sampled, to represent a certain population. The sampling strategy involves selecting the \textbf{demographics} and \textbf{number of participants} of an experiment, and considering risks related to \textbf{biases and representativeness} of those selected.

Indie game companies use participants from a diverse set of \textbf{demographics}, typically depending on the current stage of game development and the availability of participants. Our interviewees specifically distinguish between experimenting with their \textit{internal team}, \textit{friends and family}, \textit{publishers}, \textit{external unknowns}, \textit{dedicated communities}, and \textit{streamers} (i.e., individuals recording and sending live while they play and comment the game simultaneously). Several interviewees (I2, I5, I6, I10) describe how the first testing usually involves the company's \textbf{internal team} to ensure that the game \textit{``feels good''} (I2). Several interviewees (e.g. I5, I6) describe performing internal experimentation on a weekly basis, e.g. through \textit{show-and-tell sessions}. In a second step, indie game companies generally move towards including \textbf{friends and family}, e.g., through online observations or physical playtesting sessions. For example, I1 and I2 describe often recruiting users from a combination of friends and colleagues within the industry and their connections. The physical playtests are commonly hosted by local incubators or co-working spaces, where co-workers, students, and developers from the local gaming community can be found. Another important source of participants are \textbf{publishers} who are kept in the loop throughout the development. Several interviewees stress the importance of including publishers in the testing when deciding on which idea to focus on, commonly at the beginning of the pre-release stage (I2, I3, I6). As development progresses, interviewees commonly turn to \textbf{external unknowns}, i.e. to external participants in general. These are commonly recruited from indie game platforms such as Itch.io and Steam. Indie game companies publish early game versions, or demos, on these platforms to attract attention and invite players into their Discord channels (a chat platform). The Discord channels serve as an infrastructure for growing and facilitating \textbf{dedicated communities} of fans and potential playtesters that can be used in experiments on a reoccurring basis. External playtesters can also be acquired monetarily from third party platforms where the company can define the desired characteristics of the participants, to better ensure that the correct audience is reached. Another means of gaining inputs from an external audience is through the use of \textbf{streamers}. Reactions from the streamers and their audience provide valuable insights as described by I3 and I5.

The \textbf{number of participants} in an experiment varies and depends on several factors, including the type of experiment (goals, design strategy, and experiment object), type of participants, and the current stage of the game development. For physical playtests, interviewees generally converge and agree that 8-10 individuals is satisfactory. For the online and external playtests with players from the community or unknown externals, the number of participants can range from a few up to 60-70. For example, I3 indicated that around 60-70 people could be involved in an A/B test within their dedicated community. Another interviewee said that \textit{``10 players [will] often get you... [a] visible micro trend''} (I1). In contrast, I7 stated that they require a minimum of 100 testers for the input to be significant and trustworthy. However, the other interviewees explained that it is difficult to recruit that many testers. In particular, I10 highlighted the difficulty in attracting playtesters as a major challenge for their company. Generally, our interviewees described that as development progresses towards release of a game, the type and extent of experimentation shifts towards evaluation using more quantitative data from a larger number of participants.

Several interviewees described challenges in attaining \textbf{un-biased} participants (I2, I3, I4, I6) with a good \textbf{representativeness} of the target audience (I6, I7, I9). For example, friends and family often have a positive bias and are pre-disposed to approve of the company's work. Several interviewees (I1, I8, I9) mentioned that game developers provide \textit{``feedback [that is] even more useful ... [since] they have some development knowledge''} (I1). Also, game developers tend to \textit{``see things that regular gamers do not think about''} (I2).

\subsection{Execution strategy}




Feedback from those participating in an experiment is collected through a multitude of \textbf{data collection techniques}, as exemplified by the interviewees. All interviewees describe using \textbf{observations} of participants performing external game play testing. While one interviewee described the value of observing people's facial expressions (I3), most of the interviewees expressed a preference for observing \textit{``not their faces, but [to] record the screen and [to see] how they play the game''} (I2). Such recordings are \textit{``worth gold''} (I2) since they can give insights into how the game is played, and if participants get stuck or do not understand certain aspects of the game. Watching people play is described as being especially useful in mitigating biases, for example, in friends and family, or when game players know that they are talking to the developer and are, thus, less likely to criticise. Also, several interviewees described the importance of not interfering with the participants when doing live observations by asking leading questions or helping them, since this may influence their responses (I1, I3, I4). 


Several interviewees (I1-I4) described using \textbf{follow-up surveys} to complement observations. The surveys mainly consist of specific questions that the game developers want answers to, e.g. on game mechanics (if this is the focus of the experimentation), but also open questions. I2 often asks specific questions such as \textit{``did you do this? Dig there, press that button?''}, since this indicates that the gamer has understood the game logic. 

Sometimes open \textbf{follow-up interviews} are used where participants are asked to described their experience after playing the game. However, this \textit{``takes up too much time for ... sole developers''} (I2) and is an aspect for which the cost-benefit balance is a constant challenge. I6 described how experimenting with external participants takes time from development, but reduces the risk of \textit{``running into problem that are very hard or costly to correct''} (I6).

One interviewee suggested that \textbf{in-game prompts} and (existing) plugins for enabling in-game feedback could be used to gather feedback from players. For example, by inserting pops-up in the game when users \textit{``encounter issues or want to suggest something''} (I1). Such popups can be used to take screenshots and record the gamer's description of a situation. Such a feature may potentially provide valuable feedback to game developers. 

The potential and challenges of collecting and analysing large amounts of in-game \textbf{telemetry data} was raised by several interviewees (I1, I6, I7, I10). I6 uses analytics during play testing to measure, e.g. duration and progress of game play. Such information can inform about game play, e.g. difficulties, engagement, bugs etc. However, it is also \textit{``very difficult to know in advance what data you want''} (I6), which is a requirement in GDPR in order to save data. In the mobile context, I7 states that many events in these games are tightly connected to analytics, e.g. to record time from one game action to another to tell how quickly a player progresses compared to an average user. I10 reports using Google Analytics to get certain metrics, e.g., the number of active users, new players, monthly active users, day-one retention, and day-seven retention. While these general KPIs are considered helpful, I10 would prefer more detailed in-game metrics, but this would \textit{``require us to implement more analytics tooling''}. In addition to requiring technical implementation of data pipelines and processing infrastructure of the back-end, time and resources are required of the developers to collect, manage, and analyse the large amounts of data produced (I1).

\textbf{External platforms} and tools can also be used for experimentation, by disseminating games, gathering playtesters, and simultaneously collecting feedback. This data collection technique was described by I5. In this case, a prototype, or a small demo of a game was created and published as a preview on Steam to poll interest. Releasing an early access version on an external platform can provide valuable \textit{``feedback from people who explore the game'... [but also yield] very divided feedback from enthusiastic to very disappointed [gamers]''} (I5). Analysing and deciding how to address such divided feedback can be very challenging. Another common publication platform referred to is Itch.io. I9 mentioned an issues related to the limited amount of metrics this platform provides, beyond number of downloads, is used to assess the experienced fun and popularity of a game.

\section{Discussion and Conclusions}
We find that indie game development involves a continuous experimentation approach to explore, design, and develop new game ideas and features. Our interviews confirm previous research, and provide new insights into the experimentation practices used in early stages of indie game development, prior to release. In this paper, we describe these practices for five key parts of experimentation, namely goal definition, design strategy, experiment object, sampling strategy, and execution strategy, see Fig.~\ref{fig:model}.

Experimentation is commonly used in the pre-release stages for the purpose, and with the \textbf{goal} of \textit{scoping the game idea and features} for different aspects of the game. In particular, the \textit{fun factor} is characterised as a complex and difficult aspect to evaluate. Our findings align well with extant work that describes the aspects of \textit{game mechanics}, \textit{aesthetics}, and \textit{fun factor}~\cite{jarvi2015lean}. In addition, we identify the cognitive and pedagogic aspect of \textit{understandability} of the game logic as an important aspect to evaluate for more complex games. 

We note that the \textbf{design strategy}, and the formality and structure of experimentation practices, vary depending on the targeted platform (desktop vs. mobile), the level of experience and resources available to the indie company. \textit{Experimentation methods} range on a spectrum from more or less structured \textit{split (A/B) testing} (where two alternatives are evaluated in parallel) to \textit{sequential testing} (where the solution with the highest potential is evaluated) and \textit{exploratory testing} (without any pre-defined hypotheses). In contrast to earlier work~\cite{hyrynsalmi2018minimum}, our findings suggest that split testing is often performed in the initial idea stage and in early prototyping stages. In these early development stages, \textbf{experiment objects} at the \textit{conceptual level} are used predominantly. The higher cost and complexity of developing and evaluating different versions are the main reasons for limiting the split-testing to early development stages. 

The \textbf{sampling strategy} used by indie companies is greatly affected by their limited access to participants for experimentation, which presents a major challenge. However, our interview survey also reveals that in the early stages of development, indie game developers gain useful feedback from smaller sets of participants through qualitative data collection methods, such as observations and recordings of game play. In addition to obtain sufficient numbers of participant, indie companies need to consider the representatives of the target audience and the risk of obtaining biased answers, e.g., from friends and family, or monetarily awarded participants. In alignment with extant work, we also find that internal testing is commonly performed in combination with playtesting with friends and family. This risks steering the development in a direction deviating from the actual target audience~\cite{yaman2018continuous}.

 Indie game developers commonly experiment in the early development stages using an \textbf{execution strategy} that involves \textit{data collection techniques} based on qualitative data such as \textit{observations} of uninterrupted players' reactions and facial expressions. Our interviewees indicate that this is due to the limited number of users available, and is in line with other research by Yaman et al.~\cite{yaman2018continuous}. The use of quantitative data collection methods were described primarily by interviewees with experience of mobile games, which aligns with Koskenvoima and Mäntymäki~\cite{koskenvoima2015small}. However, we observed a general interest in leveraging \textit{telemetry data} to assess player engagement and understandability, e.g. by measuring the time spent on certain tasks or between two points in a game. Currently, the cost and complexity of acquiring and integrating the tools necessary to collect, manage, and analyse such data, is a barrier for small indie companies in extending their use of telemetry data in experimentation. Enabling and collecting quality feedback from the players is also reported as time consuming.

Our results may be of interest to many software-intensive organisations, irrespective of industry, and can provide insights into how experimentation can be conducted also in the early development stages when data is limited. Despite many challenges, the surveyed indie companies show high promise and resourcefulness, in particular considering the overarching high pace and competitiveness of the game industry\cite{schmalz2014risk}. Our qualitative investigation is limited to a small set of interviews of indie game developers from different studios, of which many are geographically located in Sweden. Any generalisation should be anecdotal considering the characteristics of the indie game companies surveyed and the contexts provided in this article. 

In future work, the key parts of the experimentation practices and factors influencing these will be explored and formalised further. Specific attention needs to be paid to the resource and efficiency requirements of the indie game companies, concerning tools and processes that need to be developed. Another avenue regards the exploration of perspectives and practices among more established and resourceful game companies, and how these contrast against findings of this study. An additional avenue regards the cost and benefit of introducing different practices highlighted in the proposed framework, which could guide practitioners with resource constrains as is common for indie-game developers. Further, while the focus of this study has been on the pre-release process of game development, future work should investigate how CE can be applied post-release as well.

\bibliographystyle{splncs04}
\bibliography{mybibliography}

\begin{thebibliography}{10}
\providecommand{\url}[1]{\texttt{#1}}
\providecommand{\urlprefix}{URL }
\providecommand{\doi}[1]{https://doi.org/#1}

\bibitem{aleem2016game}
Aleem, S., Capretz, L.F., Ahmed, F.: Game development software engineering process life cycle: a systematic review. J of Softw Engin Res and Dev  \textbf{4},  1--30 (2016)

\bibitem{andersen2011on}
Andersen, E., Liu, Y.E., Snider, R., Szeto, R., Cooper, S., Popovi\'{c}, Z.: On the harmfulness of secondary game objectives. In: Proc. of the 6th Int. Conf. on Foundations of Digital Games. p. 30–37. FDG '11, ACM (2011)

\bibitem{auer2021controlled}
Auer, F., Ros, R., Kaltenbrunner, L., Runeson, P., Felderer, M.: Controlled experimentation in continuous experimentation: Knowledge and challenges. Information and Software Technology  \textbf{134},  106551 (2021)

\bibitem{bjarnason2023empirically}
Bjarnason, E., Lang, F., Mj{\"o}berg, A.: An empirically based model of software prototyping: a mapping study and a multi-case study. Emp Softw Eng  \textbf{28}(5) (2023)

\bibitem{chueca2023consolidation}
Chueca, J., Ver{\'o}n, J., Font, J., P{\'e}rez, F., Cetina, C.: The consolidation of game software engineering: A systematic literature review of software engineering for industry-scale computer games. Inf. and Softw. Technology p. 107330 (2023)

\bibitem{edison2023experimentation}
Edison, H., Melegati, J., Bjarnason, E.: Experimentation in early-stage video game startups: Practices and challenges. In: Int. Conf. on Softw. Business. pp. 360--366. Springer (2023)

\bibitem{engstrom2020game}
Engstr{\"o}m, H.: Game development research (2020)

\bibitem{fagerholm2017right}
Fagerholm, F., Guinea, A.S., M{\"a}enp{\"a}{\"a}, H., M{\"u}nch, J.: The right model for continuous experimentation. Journal of Systems and Software  \textbf{123},  292--305 (2017)

\bibitem{grabarczyk2016every}
Grabarczyk, P.: Is every indie game independent? towards the concept of independent game. Game Studies  \textbf{16}(1) (2016)

\bibitem{holmstromolsson2014hypex}
{Holmstr{\"o}m Olsson}, H., Bosch, J.: The hypex model: from opinions to data-driven software development. In: Continuous software engineering, pp. 155--164. Springer International Publishing (2014). \doi{10.1007/978-3-319-11283-1\_13}

\bibitem{hyrynsalmi2018minimum}
Hyrynsalmi, S., Klotins, E., Unterkalmsteiner, M., Gorschek, T., Tripathi, N., Pompermaier, L.B., Prikladnicki, R.: What is a minimum viable (video) game? towards a research agenda. In: 17th IFIP WG 6.11 Conf. on e-Business, e-Services, and e-Society. pp. 217--231. Springer (2018)

\bibitem{swedishgamesindustry2023developerindex}
Industry, S.G.: Swedish Games Industry 2023 Game Developer Index. Swedish Games Industry (2023)

\bibitem{jarvi2015lean}
J{\"a}rvi, A., Taajamaa, V., Hyrynsalmi, S.: Lean software startup--an experience report from an entrepreneurial software business course. In: Software Business: 6th Int. Conf., 2015, Proc. 6. pp. 230--244. Springer (2015)

\bibitem{koskenvoima2015small}
Koskenvoima, A., M{\"a}ntym{\"a}ki, M.: Why do small and medium-size freemium game developers use game analytics? In: 14th IFIP WG 6.11 Conf. on e-Business, e-Services, and e-Society. pp. 326--337. Springer (2015)

\bibitem{SupplementaryMaterial}
Linåker, J., Bjarnason, E., Fagerholm, F.: Online supplementary material (2024), \url{https://doi.org/10.6084/m9.figshare.26934910}

\bibitem{pagulayan2018applied}
Pagulayan, R.J., Gunn, D.V., Hagen, J.R., Hendersen, D.J., Kelley, T.A., Phillips, B.C., Guajardo, J., Nichols, T.A.: Applied user research in games. The Wiley Handbook of Human Computer Interaction  \textbf{1},  299--346 (2018)

\bibitem{MethodRef}
Ralph, P.: Acm sigsoft empirical standards for software engineering research. qualitative surveys. arxiv:2010.03525 [cs.se] (2021), \url{https://www2.sigsoft.org/EmpiricalStandards/docs/standards?standard=QualitativeSurveys#}

\bibitem{ros2018continuous}
Ros, R., Runeson, P.: Continuous experimentation and a/b testing: A mapping study. In: Proc. 4th RCoSE. pp. 35--41 (2018)

\bibitem{rosenfield2017mvps}
Rosenfield~Boeira, J.N., Rosenfield~Boeira, J.N.: Mvps: Do we really need them? Lean Game Dev: Apply Lean Framew to the Proc of Game Dev pp. 33--48 (2017)

\bibitem{saldana2021coding}
Salda{\~n}a, J.: The coding manual for qualitative researchers. Sage (2021)

\bibitem{schmalz2014risk}
Schmalz, M., Finn, A., Taylor, H.: Risk management in video game development projects. In: 47th HICSS. pp. 4325--4334. IEEE (2014)

\bibitem{stahlke2018usertesting}
Stahlke, S.N., Mirza-Babaei, P.: Usertesting without the user: Opportunities and challenges of an ai-driven approach in games user research. Comp Ent  \textbf{16}(2) (2018)

\bibitem{su2021gameanalytics}
Su, Y., Backlund, P., Engström, H.: {Comprehensive review and classification of game analytics}. Service Oriented Comp and Appl  \textbf{15}(2),  141--156 (2021)

\bibitem{yaman2018continuous}
Yaman, S., Mikkonen, T., Suomela, R.: Continuous experimentation in mobile game development. In: Proc. of 44th SEAA. pp. 345--352 (2018)

\end{thebibliography}

\end{document}